\documentclass[5p,authoryear,times]{elsarticle}

\usepackage[modulo]{lineno}
\usepackage[]{hyperref}

\hypersetup{
	colorlinks=true,
	citecolor=red,
	linkcolor=cyan,
	filecolor=magenta,      
	urlcolor=cyan,
}

\journal{Journal of King Saud University - Engineering Sciences}

\biboptions{authoryear}

\begin{document}

\begin{frontmatter}

\title{High Impedance Fault Detection and Isolation in Power Distribution Networks using Support Vector Machines}

\author{Muhammad~Sarwar$^*$, Faisal~Mehmood, Muhammad~Abid, Abdul~Qayyum~Khan, \\ Sufi~Tabassum~Gul, Adil~Sarwar~Khan}
\address{Department of Electrical Engineering, \\
	Pakistan Institute of Engineering and Applied Sciences\\
	Islamabad, Pakistan}



%

\begin{abstract}
	This paper proposes an accurate High Impedance Fault (HIF) detection and isolation scheme in a power distribution network. The proposed schemes utilize the data available from voltage and current sensors. The technique employs multiple algorithms consisting of Principal Component Analysis, Fisher Discriminant Analysis, Binary and Multiclass Support Vector Machine for detection and identification of the high impedance fault. These data driven techniques have been tested on IEEE 13-node distribution network for detection and identification of high impedance faults with broken and unbroken conductor. Further, the robustness of machine learning techniques has also been analysed by examining their performance with variation in loads for different faults. Simulation results for different faults at various locations have shown that proposed methods are fast and accurate in diagnosing high impedance faults.  Multiclass Support Vector Machine gives the best result to detect and locate High Impedance Fault accurately. It ensures reliability, security and dependability of the distribution network.
\end{abstract}


\end{frontmatter}


\section{Introduction}

Detection of high impedance faults poses a highly challenging problem because of the random, asymmetric and nonlinear nature of high impedance fault (HIF) current. Most of the time, these faults cannot be detected and isolated by conventional over-current schemes, because magnitude of fault current is considerably lower than nominal load current \citep{233519:5217212}.

High impedance faults typically occur when an energised conductor comes in contact with ground through any high impe-dance object such as dry asphalt, wet sand, dry grass and sod etc. which limits the flow of current towards ground \citep{233519:5217212}. Timely detection of high impedance faults is necessary for efficient, reliable and safe operation of power systems. Probability of occurrence of high impedance fault in distribution networks is more than in transmission network because distribution feeders are more likely to come in contact with high impedance objects like trees etc. However, in underground cables high impedance faults are caused by insulation degradation that exposes the energised conductor to high impedance objects \citep{233519:5217213}. High impedance faults occur at voltage level of 15KV or below in most of the cases. Magnitude of HIF current is independent of the conventional short circuit fault current level \citep{233519:5217214}. 

High impedance faults are extremely difficult to detect and isolate by conventional protection schemes, because fault current magnitude is much lower than nominal current. According to report by Power System Relaying Committee (PSRC), only 17\% of HIFs can be detected by conventional relaying schemes \citep{233519:5217212}.
Detection of HIFs helps in prognostic maintenance in power distribution system. High impedance faults involve arcing which makes fault current asymmetric and nonlinear. As a result of arcing, HIFs involve high frequency components similar to load and capacitor switching which makes detection much more difficult \citep{233519:5217215}.

Previous research on diagnosis of HIFs was focused on lab-based staged fault studies. However, with the advancement in technology and better understanding of features of HIFs, the focus has shifted towards simulations and software studies \citep{233519:5217216}.


Due to its critical nature, researchers from both industry and academia have proposed various techniques to detect HIFs in distribution networks. Majority of the studies were reported as early as 1980s and 1990s, but the simulation methods and advanced detection techniques are still being developed and proposed. HIF detection methods can be broadly classified into time domain algorithms, frequency domain algorithms \citep{lima2018high}, hybrid algorithms \citep{samantaray2008high} and knowledge-based systems \citep{etemadi2008high}.

K. Zoric and M. B. Djuric presented a method to detect high impedance fault based on harmonic analysis of voltage signals \citep{233519:5217219}. James Stoupis introduced a new relaying scheme manufactured by ABB in the area of artificial neural networks \citep{233519:5217220}. Sedighi proposed two methods based on soft computing for detection of HIF \citep{233519:5217221}.
Mark Adamiak proposed signature based high impedance fault diagnoses which involves expert system pattern recognition on harmonic energy levels in arcing current \citep{233519:5217222}.

S.R. Samantaray presented an intelligent approach to detect high impedance faults in distribution systems \citep{233519:5217224}. In \citep{233519:5217225}, authors have proposed a new approach to detect HIF based on PMU (Phasor Measurement Unit).
F. V. Lopes presented a method to diagnose HIF in smart distribution systems \citep{233519:5217226}. Authors in \citep{233519:5217216,233519:5217223} presented a method to detect HIF based on mathematical morphology. A new model for high impedance faults has been present in \citep{233519:5217227}. Results of this model are quite closer to what is observed in staged faults. This research activity also detected HIF using harmonic analysis of current waveform.

Recenlty, Kavi presented a method to detect HIF in Single Wire Earth Return (SEWR) system \citep{233519:5217228}. Sekar and Mohanty proposed fuzzy rule base approach for high impe-dance fault detection in distribution systems \citep{sekar2018fuzzy}. They present a filter-based morphology gradient (MG) to differentiate non-HIF events from HIF events.
W. C. Santos presented a transient based approach to identify HIF in power distribution systems \citep{233519:5217230} . In \citep{233519:5217231}, real time complexity measurement (RCM) based approach is used to detect HIF. In \citep{233519:5217217} HIF detection techniques are evaluated and compared with each.
With the advancement in technology, the trend has been shifted towards smart grids and smart distribution systems. Smart distribution systems include measurements (such as voltage and current) at each node that helped to discover and develop digital signal processing based fault detection techniques \citep{233519:5217232,233519:5217233,233519:5217234}.

Prior studies have helped to reveal many of the hidden characteristics of High Impedance Faults. But the major drawback of aforementioned techniques is that they are not capable of detecting \textit{all} types of high impedance faults. Furthermore, active methods of HIF detection use signal injection which deteriorates the power quality. Some methods employ data gathered by PMUs, which are quite expensive and also many distribution systems currently don't deploy PMUs. Another disadvantage is that most of the proposed methods use a lot of computing power and thus can not be implemented on an embedded system as a portable numerical relay.

The motivation for this research work lies in manifold shortcomings of the prior research. The proposed method is computationally less rigorous, so it can be implemented on an embedded system as a numerical relay. The method ensures power quality as it does not inject any signal into power system for HIF detection. The technique developed can detect all types of HIF i.e., broken and unbroken conductor HIFs and can locate the faulty section of network. Furthermore, input data is gathered using CTs and PTs which are already deployed in all distribution networks, so no additional hardware installation is needed for data acquisition. The proposed method is accurate and highly reliable as it can distinguish load switching from faults, and can also detect and isolate multiple high impedance faults in the power network. The training model used for detecting faults is of low order and can easily be implemented on any embedded hardware for real time prototyping and HIF detection.

In proposed method, data obtained from voltage and current sensors is fed to three fault detection algorithms i.e., Principal Components Analysis (PCA), Fischer Discriminant Analysis (FDA) and Support Vector Machines (SVM). PCA extracts principal components of data and use them for fault detection. FDA reduces the data to a lower dimension to maximise distance among various classes for increased accuracy. Binary class SVM is used only for detection of HIFs. To determine type and location of fault multiclass SVM is deployed to signal the presence of an incipient or sudden high impedance fault. Once the fault is detected, the faulty system can be isolated from the network by issuing a trip signal to traditional over current relays in substation. The speed and accuracy of proposed method is comparable to conventional fault detection of overcurrent faults.

The rest of research paper is organised as follows. Section II details characteristics and features of high impedance faults. Theoretical foundation for data driven techniques is laid down in Section III. Proposed techniques have been tested on IEEE 13-node test system and results are discussed in Section IV. Section V concludes the research.

\section{High Impedance Fault Characteristics}
In this section, prominent features of high impedance faults are described. To obtain training data from simulation, a fault model is simulated in Simulink and training data is obtained from voltage and current sensors installed in the network.

\subsection{Properties of High Impedance Faults} 
Arcing is a prominent phenomena in HIFs. The arc is formed due to air gap between energised conductor and high impedance object. Arc ignition occurs when magnitude of voltage is higher than air gap breakdown voltage. Consequently, arc extinction occurs when voltage is lower than breakdown voltage \citep{233519:5217217}. The value of break down voltage changes during each cycle. Thus, in every cycle of voltage, the HIF current includes two arc re-ignitions and two arc extinctions. Therefore, current conducting path changes during each cycle which changes the magnitude of HIF current making it non-linear, also HIF current is intermittent in nature \citep{233519:5217217}. Some of the typical features of high impedance faults are as follows:
\begin{itemize}
	\item \textbf{Non linearity:} Voltage-current characteristics are highly nonlinear due to change in current conducting path \citep{233519:5217238}.
	\item \textbf{Asymmetric nature of HIF current:} Peak values of current are different in positive and negative half cycle due to the presence of varying break down voltage \citep{233519:5217217,233519:5217235}.
	\item \textbf{Intermittent nature:}  HIF current is not steady due to intermittent nature of arc \citep{233519:5217236}.
	\item \textbf{Build up:} Current magnitude progressively increases till it reaches it maximum value \citep{233519:5217235}
	\item \textbf{Randomness:} Magnitude of HIF current and its shape changes with time due change in impedance of conducting path \citep{233519:5217237}.
	\item \textbf{Low and high frequency components:} HIF current includes low frequency components due to non-linearity of HIF. Additionally, HIF current also contains high frequency components due to intermittent nature of arc\\ \citep{233519:5217217}.
\end{itemize}

\subsection{Simulation of High Impedance Fault} 
\begin{figure}
	\centering 
	\includegraphics[height=6cm]{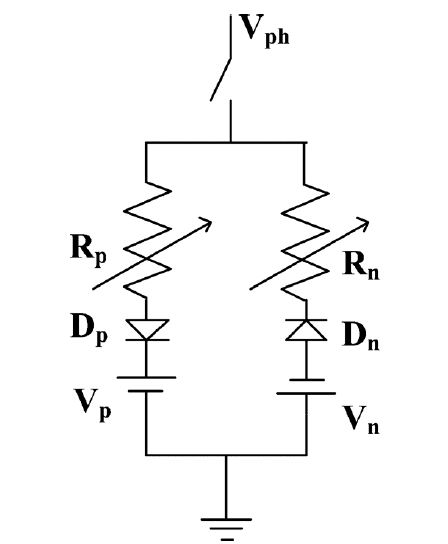}
	\caption{{The selected HIF model \citep{233519:5217216}}}
	\label{hifmodel}
\end{figure}

To obtain training data from simulation, appropriate model of HIF is required which would show real behaviour of an HIF. This paper utilises an HIF model shown in Fig.~\ref{hifmodel}, connected between any of phase and ground \citep{233519:5217216}. The model is simulated in MATLAB, the parameters of model are tuned according to test feeder. 

In this HIF model, two diodes $D_p $ and $D_n $ are connected to two DC voltage sources $V_p $ and  $V_n$, respectively. The DC sources have different magnitude and their magnitude randomly changes around  $V_p$ and $V_n $ after every 0.11 ms. This models the asymmetric nature of arc current and intermediate arc extinction. 

To model the randomness in duration of arc extinction in high impedance fault, voltage polarity also changes at every sampling instant \citep{233519:5217216}. When the instantaneous value of phase voltage is greater than  $V_p $, current flows towards ground, when the instantaneous value of phase voltage is less than $V_n $, current reserves its direction, when the instantaneous value of phase voltage is between  $V_p $, and  $V_n $, no current flows. In order to incorporate varying arc resistance, the model of HIF also includes two variable resistances, $R_p $ and  $R_n $, such that values of these resistances vary randomly after every 0.11 ms.

The parameters used for HIF model with IEEE 13-node test feeder in Simulink are:

\begin{eqnarray*}
	\begin{array}{l}\begin{array}{l}V_p=1.0\;kv,\;with\pm\;10\%\;variation\\V_n=0.5\;kv,\;with\pm\;10\%\;variation\end{array}\\R_p,R_n=1000~\Omega-1500~\Omega,\;with\;random\;variation\end{array}
\end{eqnarray*}

The above model is simulated in Matlab{\textregistered}. There are two steps involved in modelling HIF; in first step, variable DC voltage sources are modelled using controlled voltages source; in second step, variable resistances are modelled using controlled current sources. First part is implemented using only random number generator, constant block, and controlled voltage source is used to obtain varying DC voltages. Second step involves a build-up series R-L circuit and a sinusoidal signal of 60 Hz. Both of these generate an exponentially growing sine wave. The sine wave is multiplied with a random number of amplitude 1 and variance of 0.12 to obtain a randomly varying resistance.

\begin{figure}[tb]
	\centering \includegraphics[width=0.45\textwidth]{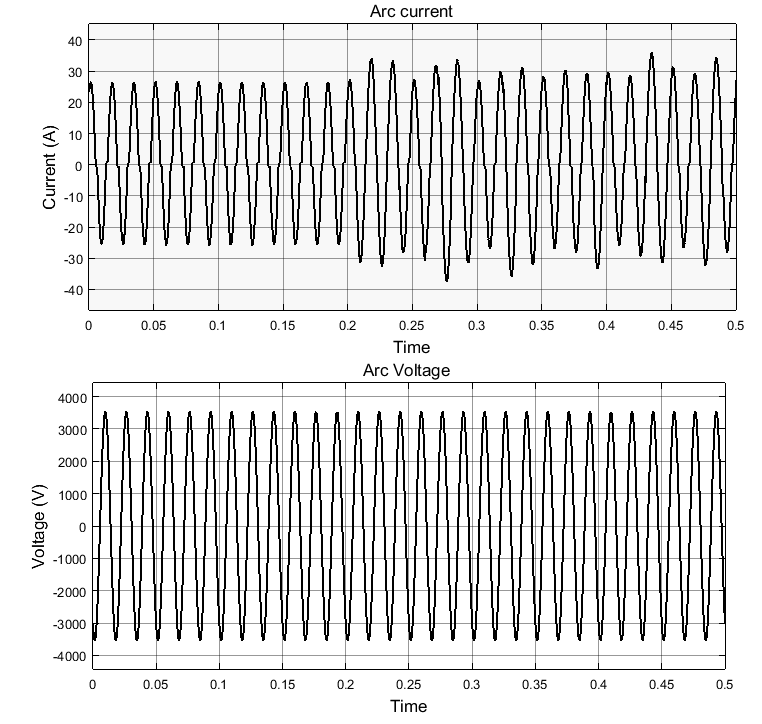}
	\caption{{Arc current and voltage during HIF}}
	\label{viHIF}
\end{figure}



\begin{figure}[tb]
	\centering \includegraphics[width=0.45\textwidth]{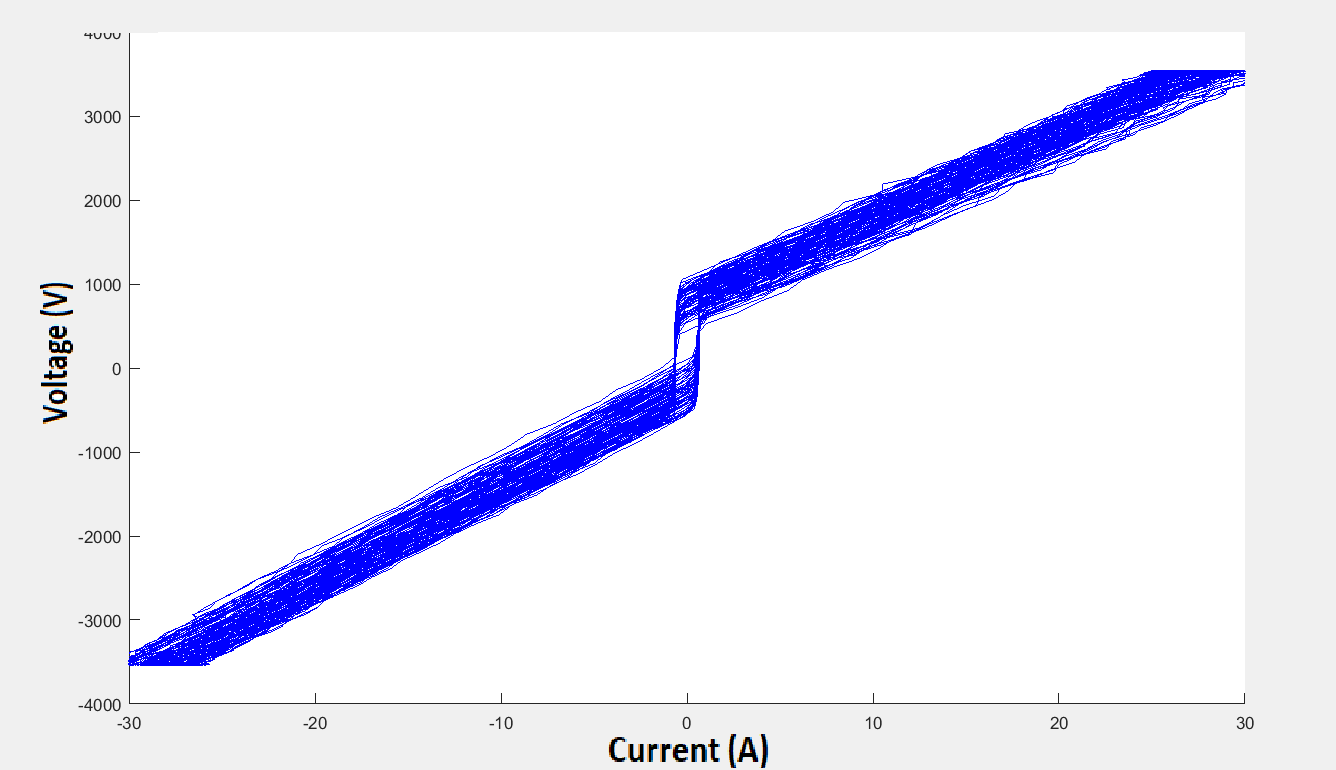}
	\caption{{The v-i characteristics of HIF}}
	\label{viChar}
\end{figure}

Fig.~\ref{viHIF} shows arc current and voltage waveforms obtained as a result of modelling HIF in Simulink, using HIF model of Fig.~\ref{hifmodel}. It is clear that the arc current is small, random, asymmetric, and nonlinear in nature. The voltage waveform in Fig.~\ref{viHIF}  also shows the random behaviour. Fig.~\ref{viChar} shows the v-i characteristics of HIF, the v-i characteristics of HIF and the current waveform are quite similar to those got from a staged fault \citep{233519:5217216}.

\section{Theoretical Foundation}
Data driven techniques are the perfect candidate for fault diagnoses in large systems where enough amount of data is available. Principal Component Analysis, Fischer Discriminant Analysis, and Support Vector Machines are widely used for addressing diagnosis problems due to their simplicity and efficiency in processing large amount of data. Here the theoretical basis for applied algorithms is given.

\subsection{Principal Component Analysis} 
Principal Component analysis is linear dimensional reduction technique, it projects higher dimensional data into lower dimensions while keeping significant features. PCA has ability to retain maximum variation that is possible in lower dimensions such that transformed features are linear combination of primary features. In reduced dimensions, different statistical plots such as $T^2 $ or Q-charts are utilised for visualisation of different trends. PCA is known as powerful tool for feature extraction and data reduction in fault detection techniques because of simplicity and its ability to process large amount of data \citep{233519:5217241}.

Application of PCA for fault diagnoses consists of three steps; first of all, loading vectors (transformation vectors) are  calculated by performing offline computations on training data; in second step, the loading vectors are utilised to transform online data (higher dimensional data) into lower dimensions; in third step, test statistics such as $T^2$ are used to detect fault \citep{233519:5217241}.

\begin{figure}[tb]
	\centering 
	\includegraphics[scale=0.5]{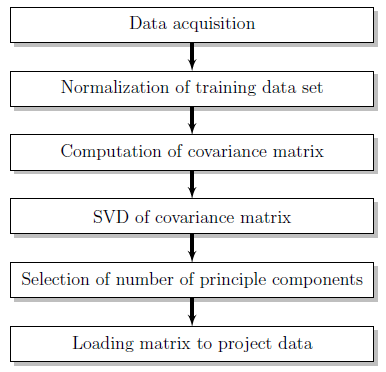}
	\caption{{Flowchart for offline fault computation using PCA}}
	\label{PCA-offline}
\end{figure}

\begin{figure}[tb]
	\centering 
	\includegraphics[width=0.45\textwidth]{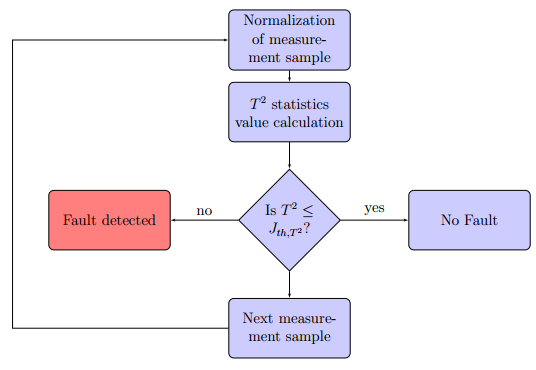}
	\caption{Flowchart for online fault computation using PCA}
	\label{PCA-online}
\end{figure}

Let us assume that a training set of $m$ process variables, with set of $n$ observations, is normalised to unit variance and zero mean by subtracting each process variable by its mean and dividing by standard deviation of data, and is shown in the form of input matrix $\mathrm{X}\mathrm{\in }{\mathrm{R}}^{\mathrm{n\times m}}$.  With the help of singular value decomposition of input data matrix X, loading vectors or transformation vectors are calculated.
\begin{eqnarray}
\frac{1}{\sqrt{1-n} } X=U{\rm \Sigma V}^{{\rm T}}
\label{eq1}
\end{eqnarray}
In equation (\ref{eq1}), $U$ and $V$ are unitary matrices, and $\Sigma $ is called diagonal matrix and its singular values are in decreasing order. 
The transformation vectors are orthonormal vectors of matrix $\mathrm V\in\mathrm R^{\mathrm m\times\mathrm m} $. Training set's variance projected along the $u^{th}$ column of V is equal to $\sigma _u^2 $. In PCA, loading vectors or transformation vectors related to a largest singular value are kept to capture large data variation in lower dimensions. Let us assume that $\mathrm P\in\mathrm R^{\mathrm m\times\mathrm a} $ is the matrix with first a column of  $\mathrm V\in\mathrm R^{\mathrm m\times\mathrm m} $, and projection of observed data X into reduced dimensions are incorporated in the score matrix T is given as,
\begin{eqnarray}T=XP\end{eqnarray}
Once the data is projected in lower dimensions, Hotteling's $T^2 $ statistics is used for fault detection. Hotteling's $T^2 $- statistics can be calculated as \citep{233519:5217241,233519:5217242},
\begin{eqnarray}
T^2=x^T P\Sigma_a^{-1} P^T x
\label{eq3}
\end{eqnarray}
where $\Sigma$ is the diagonal matrix of first $a$ singular values, $P$ is the loading vector matrix corresponding to first a singular values. The Hotteling's $T^2 $- statistics (\ref{eq3}) is scaled squared 2-norm of observation space X, measures systematic variations of the process, and if there is violations, it will indicate that systematic variations are out of control. If \ensuremath{\alpha } is the level of significance, the threshold of  $T^2 $ statistics can be calculated as \citep{233519:5217242},
\begin{eqnarray}T_\alpha^2=\frac{m(n-1)(n+1)}{n(n-m)}F_\alpha (m,n-m) \label{eq4}\end{eqnarray}
Where $F_\alpha (m,n-m) $ is known as F-distribution with m and (n-m) degree of freedom \citep{233519:5217242}. Essential condition for fault detection occurs if Hotteling's $T^2 $- statistics exceeds its threshold value, that is,
\begin{eqnarray*}T^2 \ensuremath{\leq } T_\alpha^2 \ \ \ \       Fault\ free\ case \\
	T^2 > T_\alpha^2 \ \ \ \ \ \ \  \ \ \ \ \ \ s   Fault\  case\end{eqnarray*}
A complete flowchart for offline and online computation of the PCA algorithm for fault detection is shown in Fig. \ref{PCA-offline} and \ref{PCA-online}, respectively.

\subsection{Fisher discriminant analysis}
Fisher discriminant analysis is one of the most powerful methods for dimensionality reduction. In case of fault detection, PCA gives very good results. However, it has poor properties of fault classification because it does not consider information (variance) among different classes of data during computation of loading vectors \citep{233519:5217241}. FDA considers information among different classes of data, so it is more favourable for fault classification. It determines a set of transformation vectors, known as FDA vectors. FDA vectors maximise the information (distance) among different classes of data, while minimising information within each class in projected space. FDA tries to centralise different data classes and feature recognition rates of FDA is better than PCA. According to \citep{233519:5217240}, performance of FDA for fault detection and classification is quite better than that of PCA.

The procedure to implement FDA is similar to PCA. First of all, FDA vectors are computed using training data, then these FDA vectors are utilised to transform online data into lower dimensional space. Finally, a discriminant function isolates the fault. In FDA training data, both normal and faulty data is used for computation of FDA vectors, however, in PCA only normal data is used for computation of loading vectors \citep{233519:5217240,233519:5217242}. In order to detect fault with the help of FDA, Hotteling's $T^2$- statistics is used. 

Let us assume that a training set of m process variables, with set of n observations, is shown in the form of input matrix $\mathrm{X}\mathrm{\in }{\mathrm{R}}^{\mathrm{n\times m}} $. Consider $q$ as number of classes in different faults and $n_k$ is number of observations in $k^th$  class, let $x_i $ be the transpose of \textbf{\textit{i}}\ensuremath{_{th}} row of matrix X. The transformation vector \ensuremath{\nu} is computed using training data such that following optimisation is solved.
\begin{eqnarray}
J_{FDA}(\nu)=arg \ max\ _{\nu\neq0} \frac{\nu^T  S_b \nu}{\nu^T  S_w \nu}
\label{eq5}
\end{eqnarray}
Where S\_w shows within class scatter matrix given by
\begin{eqnarray}S_w= \Sigma_{k=1}^qS_k \end{eqnarray}
With
\begin{eqnarray}S_k=\Sigma_{x_i\in\ensuremath{\chi }_k}^n(x_i-\overline{x}_k) (x_i-\overline{x}_k)^T\end{eqnarray}
and the mean of kth  class $\overline{x}_k=\frac1n_k \Sigma_{x_i\in x_k}x_i  $  similarly $S_b $ is between class scatter matrix given by
\begin{eqnarray}S_b=\Sigma_1^q(x_i-{\overline x}_k)(x_i-{\overline x}_k)^T\end{eqnarray}
With $\overline x $ shows the combined (total) mean vector given by $\overline{x}=\frac1n \Sigma_{i=1}^n \ x_i $ it is stated that solution to above optimisation problem is identical to eigenvalue decomposition problem \citep{233519:5217243},

\begin{eqnarray}
S_b \nu_h=\lambda_h S_w \nu_h
\end{eqnarray}

\begin{figure}[h]
	\centering 
	\includegraphics[width=0.5\textwidth]{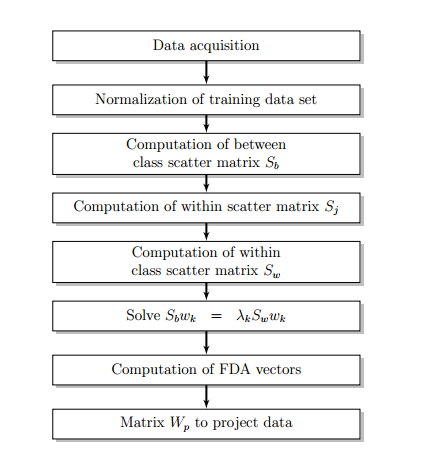}
	\caption{Flowchart for offline training of FDA}
	\label{offlineTrainFDA}
\end{figure}

Where $\lambda_h $ is generalised eigenvalue representing the extent of separability between classes and $\lambda_h $ are respective eigenvectors. Equation (\ref{eq5}) shows optimisation problem that ensures minimum scatter within class and maximum scatter between different data classes. This feature helps to classify faults.
In order to project online data into lower dimensional space, a matrix $V_q \in R^{(m\times q-1)} $   with q-1 FDA vectors is defined as, such data projected data $z_i \in R^{(q-1)} $ is given by
\begin{eqnarray}z_i = V_q^T  x_i\end{eqnarray}
For fault detection Hotteling's $T^2 $- statistics is used \citep{233519:5217244}, given by
\begin{eqnarray}T_k^2 = x^T V_a  (V_a^T S_k V_a )^{-1} V_a^T x
\end{eqnarray}
Where a shows the number of non-zero eigenvalues. For a given level of significance \ensuremath{\alpha } , threshold for Hotteling's $T^2 $- statistics is given by:
\begin{eqnarray}
T_\alpha^2= \frac{a(n-1)(n+1)}{n(n-1)}{F_\alpha (a,n-a)}  \\
T_k^2 \ensuremath{\leq } T_\alpha^2  \ \ \    Fault\ free\ case  \nonumber \\
T^2>T_\alpha^2       Fault\ case \nonumber
\end{eqnarray}
For fault classification, the discriminant function is used as given below:
\begin{eqnarray}
g_k(x)= -\frac12(x-\overline x_k)^T V_q {(\frac{1}{n_k-1} V_q^T S_k V_q})^{-1} V_q^T (x-\overline{x}_k) \nonumber\\ 
+ln(q_i )-\frac12  ln[det(\frac{1}{n_k-1} V_q^T S_k V_q )]
\end{eqnarray}
In above equation, g\_k (x) is the discriminant function associated with class k, provided a data vector $x{\in}R^m$, online data is associated with class $i$ provided that the discriminant function belonging to $i^th$ class is maximum for a fault in class $i$, can be expressed as,
\begin{equation}
g_i (x)  > g_k (x)
\end{equation}
A complete flowchart for offline training of the FDA algorithm is shown in Fig. \ref{offlineTrainFDA}.

\subsection{Support Vector Machines (SVM)}
\begin{figure}
	\centering 
	\includegraphics[width=0.35\textwidth]{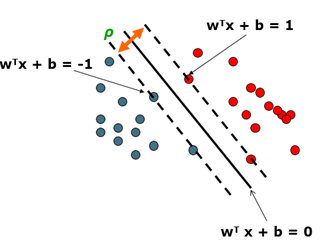}
	\makeatother 
	\caption{{Linear separating hyperplane \protect\citep{233519:5217247}}}
	\label{linePlanFig4}
\end{figure}

SVM is a well-known data driven technique used for detection and classification of faults due to its generalisation ability and being less susceptible to the curse of dimensionality \citep{233519:5217245}. For the first time, Support vector machines were used by Vapnik \citep{233519:5217246}. It is one of the new machine learning tools for classification of linear and nonlinear data. SVM is a binary classifier that maximises the margin between two data classes through a hyper-plane as shown in Fig. \ref{linePlanFig4}. SVMs maximise the margin near separating hyperplane. The decision of separation is fully identified by the support vectors. Solution of SVM is obtained through solution of quadratic programming. 

In SVM, a discriminant function is used to differentiate different classes of data given by:
\begin{eqnarray}f(x)=w^T x+b\end{eqnarray}
Where b, the bias, x, the data points, and w, the weighting vector, are obtained through training data. In two-dimensional space, the discriminant function is a line, in three-dimensional space, the discriminant function is a plane, and in n-dimensional space, the discriminant function is a hyperplane. SVM generates the optimal separating hyperplane by calculating the value of bias, weighting vector in such way that maximum margin is achieved. The points in training set with least perpendicular distance to the hyperplane are known as support vectors. The margin of the optimal separator can be defined as width of separation between support vectors.
\begin{eqnarray}\rho=2 \frac{f(x^0  )}{w} =2r\end{eqnarray}

\subsubsection{The Kernel Trick (Feature Space)}
\begin{figure}
	\centering 
	\includegraphics[width=0.5\textwidth]{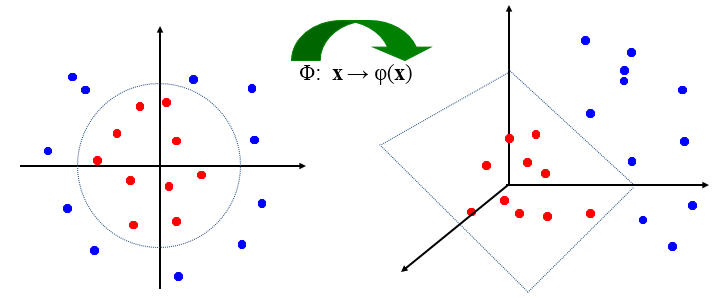}
	\makeatother 
	\caption{{Mapping of data to feature space \protect \citep{233519:5217247}}}
	\label{featureSpace}
\end{figure}

The cases in which training data is not linearly separable in the original space using above methods, then, this kind of data can be mapped to a higher-dimensional space which makes the data separable \citep{233519:5217247}, as shown in Fig.~\ref{featureSpace}.

A kernel function is a type of function that corresponds to an inner product in the higher dimensional space. For example, if data is mapped to feature space through a transformation $\Phi:\;\;x\;\rightarrow\;\varphi(x) $, then, the inner product results:
\begin{eqnarray}K(xi, xj) = \phi(xi) T \phi (xj)
\end{eqnarray}
There are different types of kernels, such as, polynomial, linear, Radial Basis Function (RBF) etc. The discriminant function of SVM, can be written as:
\begin{eqnarray}f(x)=w^T x+b\end{eqnarray}
According to Representer theorem, $w$ can be written as linear combination of input vectors.
\begin{eqnarray}w= \Sigma_{j=1}^N \alpha_j x_j\end{eqnarray}
Thus
\begin{eqnarray}f(x)=w^T x+b=b+ \Sigma_{l=1}^N \alpha_l x_l^T x\end{eqnarray}
All the dot products can be replaced with
\begin{eqnarray}k(c,d)=c^T d
\end{eqnarray}
Optimisation problem:
\begin{eqnarray}min \  a,b \ \  \frac12 \Sigma_{j,l=1}^N \alpha_j \ensuremath{\alpha }_l k(x_j,x_k )+C \Sigma_{j=1}^N \xi_j
\end{eqnarray}
Where \ensuremath{\xi }\_j{\textgreater}0
\begin{eqnarray}y_j \Sigma_{l=1}^N \alpha_l k(x_l,x_j)+b\ensuremath{\geq }1- \xi_j
\end{eqnarray}
In order to test the pattern, we use:
\begin{eqnarray}f(x)=b+ \Sigma_{l=1}^N \alpha_l k(x_l,x)\end{eqnarray}
Euclidean dot product can be substituted with dot product in feature space  ``\ensuremath{\Phi }'', which will permit nonlinear classification.
\begin{eqnarray}k(c,d)= \Phi(c)^T \Phi(d)\end{eqnarray}
$k(c,d)$ is known as kernel function and corresponding SVM is called kernelized SVM. This type of SVM can solve the issue of classification of not linear separable data. 
Steps involved in implementation of kernelized SVM are:\\

\begin{enumerate}[Step 1:]
	\item Input data is normalised.
	\item Training of SVM.
	\begin{enumerate}[Step 2.1:]
		\item Selection of kernel function.
		\item  Selection of kernel parameter.
		\item  Optimisation of penalty factor (C).
		\item  Cross validation.
	\end{enumerate}
	\item  Classification of SVM test data.
\end{enumerate}

\subsubsection{Multiclass SVM}
Binary class SVM can be used for fault detection, but it cannot be used for fault classification. However in practical cases, discrimination of more than two classes is required, hence, multiclass pattern recognition is often required in real world problems \citep{233519:5217248}. In majority of cases, multiclass pattern recognition problems are decomposed into series of binary problems such that binary pattern recognition techniques can easily be applied in practical cases. multiclass SVM algorithms such as one-versus-one, one-versus-all, can be applied be applied to classify more than two faults.

\section{Application of Data Driven Techniques to Diagnose HIF}
\begin{figure}
	\centering \includegraphics[height=6cm, width=0.45\textwidth]{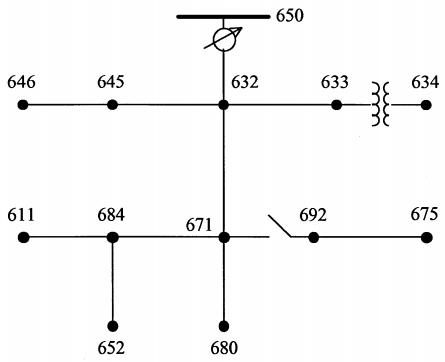}
	\makeatother 
	\caption{{IEEE 13-node Distribution Test Feeder}}
	\label{13node}
\end{figure}

HIF is introduced at different positions and different phase conductors of IEEE 13-node test feeder as shown in Fig.~\ref{13node}. In data structure, data is generated from Simulink model of test feeder. There are 29 variables of singles phase, two phase and three phase voltages of 13-node test feeder. Data has been placed in input matrix in such a way that each column of input matrix represents voltage and each row of input matrix represents number of observations. There were 400 observations recorded for bus voltages, first 100 observations correspond to normal data, while other 300 observations correspond to three HIF locations at different positions of test feeder.

\subsection{Detection of HIF using PCA and Hotteling's $T^2 $ statistics}
For diagnoses of HIF using PCA, training data consisting of 60 samples of normal condition (without fault) has been selected while testing data is consist of 100 samples of non-faulty data and 100 samples of faulty data. PCA algorithm has been applied on training data and 29 principal components are obtained. Out of 29 principal components only 5 principal components have been retained, the decision is made on the basis of total variance captured by these 5 principal components. The value of $\alpha$, as mentioned in (\ref{eq4}), is taken 0.001. As we have retained 5 principal components so $(\frac{1.6983}{1.72})=98\% $ of total variance has been captured by first five principal components.

\begin{figure}
	\centering \includegraphics[width=0.5\textwidth]{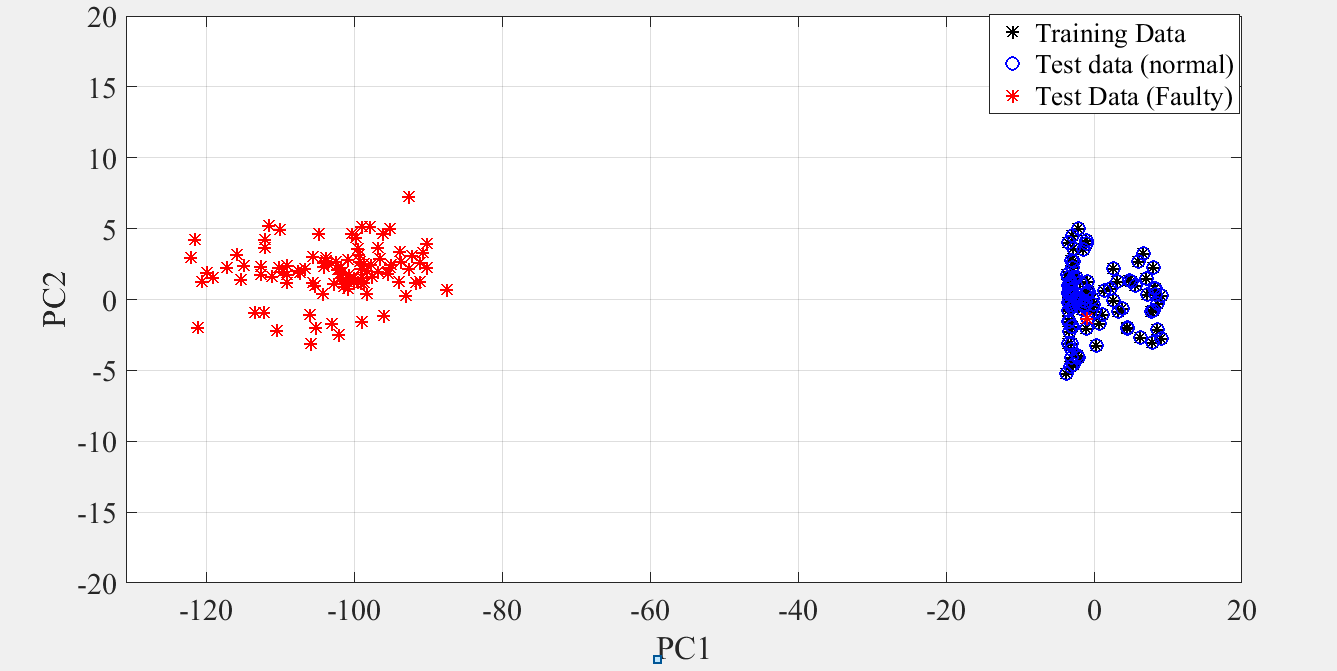}
	\makeatother 
	\caption{{Projection of training data and testing in two dimensional space for PCA}}
	\label{projectData}
\end{figure}

\begin{figure}
	\centering \includegraphics[width=0.5\textwidth]{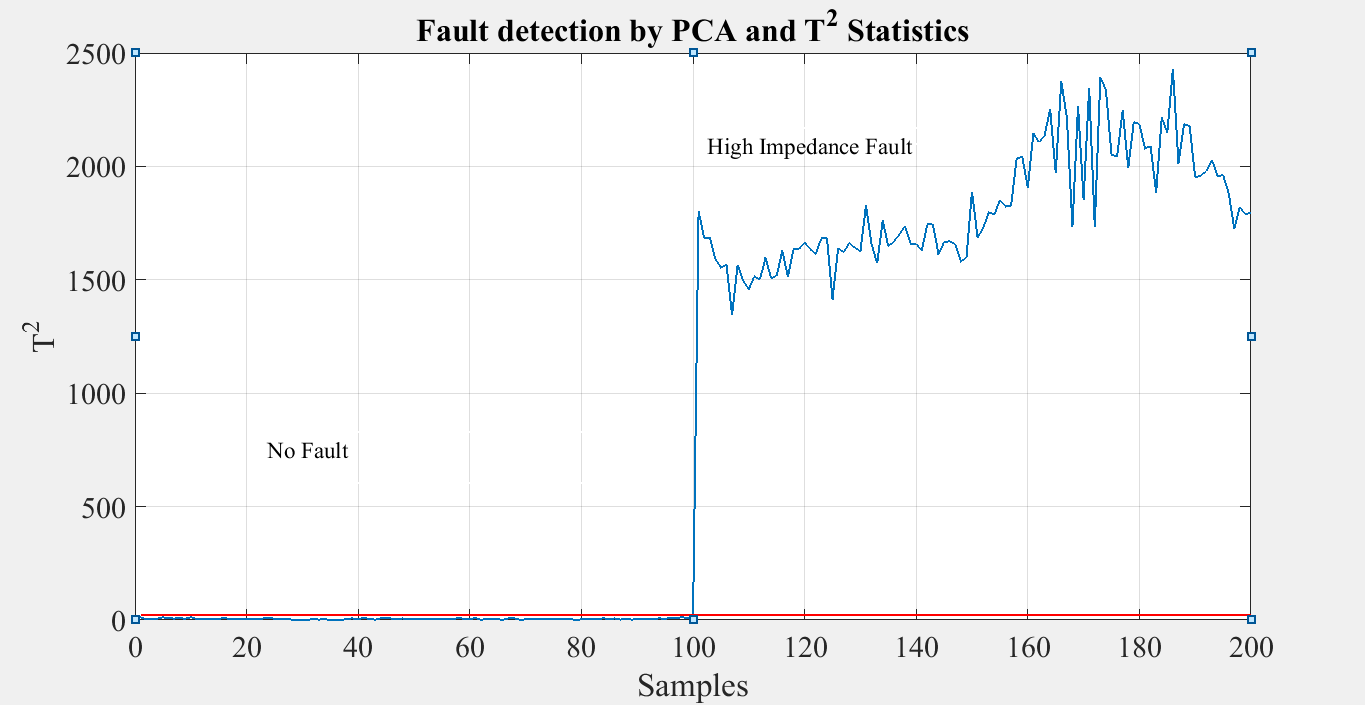}
	\makeatother 
	\caption{{The results of Hotteling's $T^2$ statistics to detect HIF}}
	\label{hotteling}
\end{figure}

Fig.~\ref{projectData} shows projection of training data and testing in two dimensional space. It can be observed that first two components capture most of variation in higher dimensional data. Fig.~\ref{hotteling}  shows the results of Hotteling's $T^2$ statistics to detect HIF after applying (3.3) on test data. 
It can be seen that normal data (first 100 samples) lies below threshold value of $T^2$ statistics, where threshold value is 22.0108. This threshold value was found using significance level of 0.1\% and confidence region of 99\%.

\begin{figure}
	\centering \includegraphics[width=0.5\textwidth]{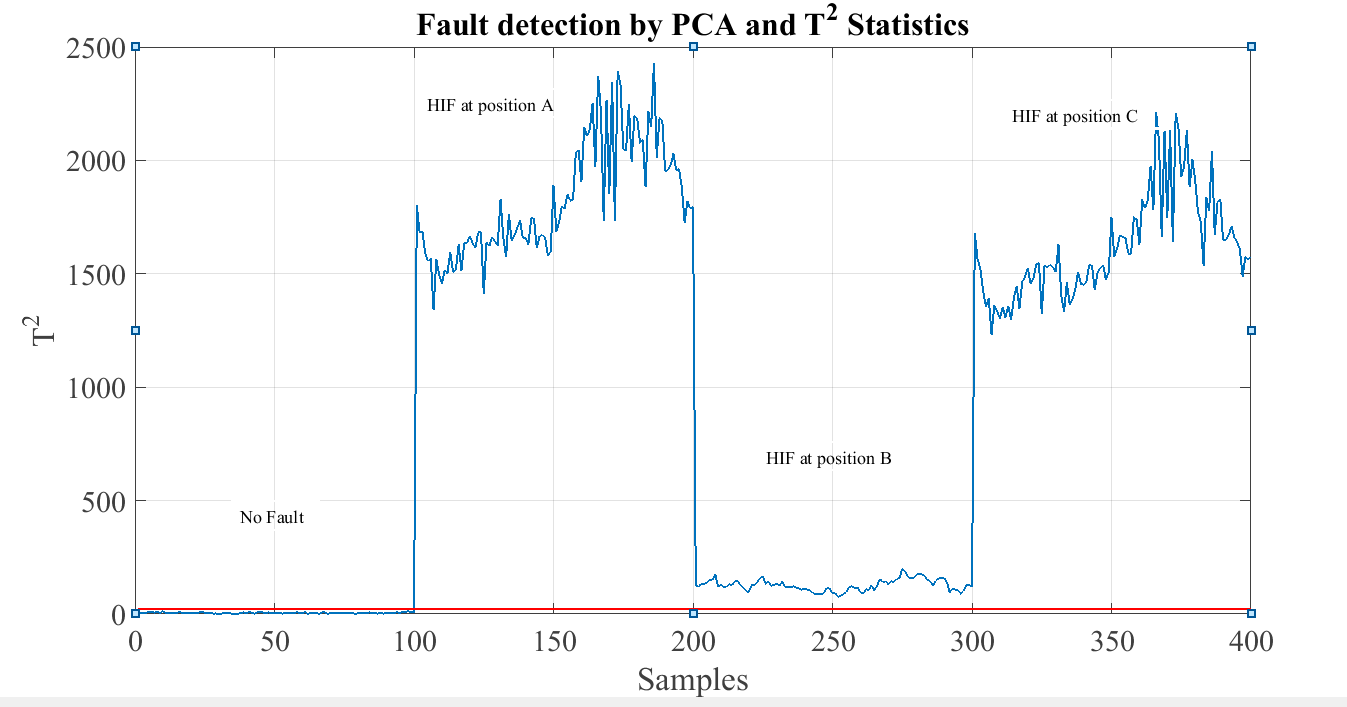}
	\makeatother 
	\caption{{Plot of Hotteling's $T^2$ statistics to detect multiple HIFs}}
	\label{hotteling3location}
\end{figure}

\begin{figure}
	\centering \includegraphics[width=0.5\textwidth]{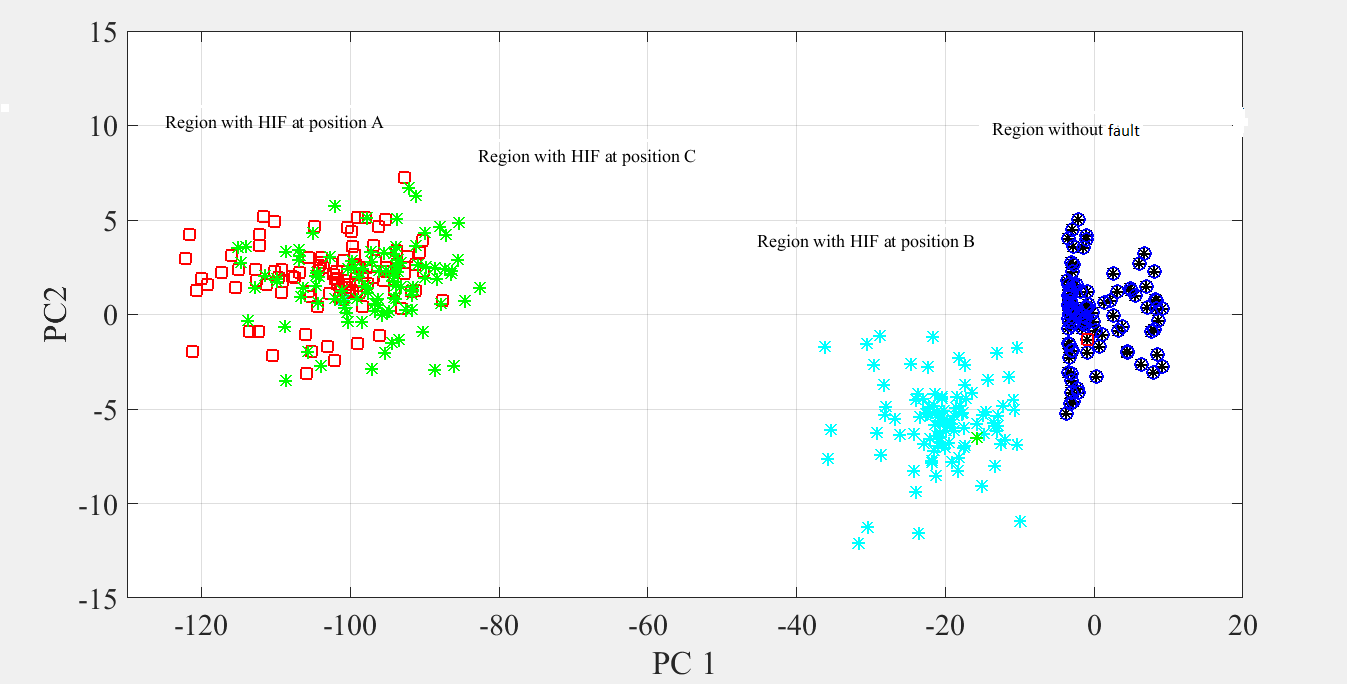}
	\makeatother 
	\caption{{Projection of training and test data in 2-D space for multiple faults}}
	\label{multiFaultproject}
\end{figure}

PCA can successfully detect high impedance fault as shown in Fig.~\ref{hotteling}. In some cases, it is required to classify different types of HIFs such as broken conductor and unbroken conductor HIFs at different locations of feeder. For this purpose, High impedance faults at three different locations are analysed. Fig.~\ref{hotteling3location} shows plot of Hotteling's $T^2 $ statistics to detect HIFs at three different locations. Results show that PCA can successfully detect these three HIFs. Only 5 principal components are retained such that  (1.6983/1.72)=98\% of total variance has been captured. Fig.~\ref{multiFaultproject} shows projection of training data and test data in two dimensional space, it can be seen that PCA cannot discriminate between different types of HIFs, this is due to the reason that PCA do not consider information among different classes of data. We can conclude that PCA is suitable for HIF detection but it cannot classify different types of HIFs.

\subsection{Detection of HIF using FDA}
\begin{figure}[tb]
	\centering \includegraphics[width=0.5\textwidth]{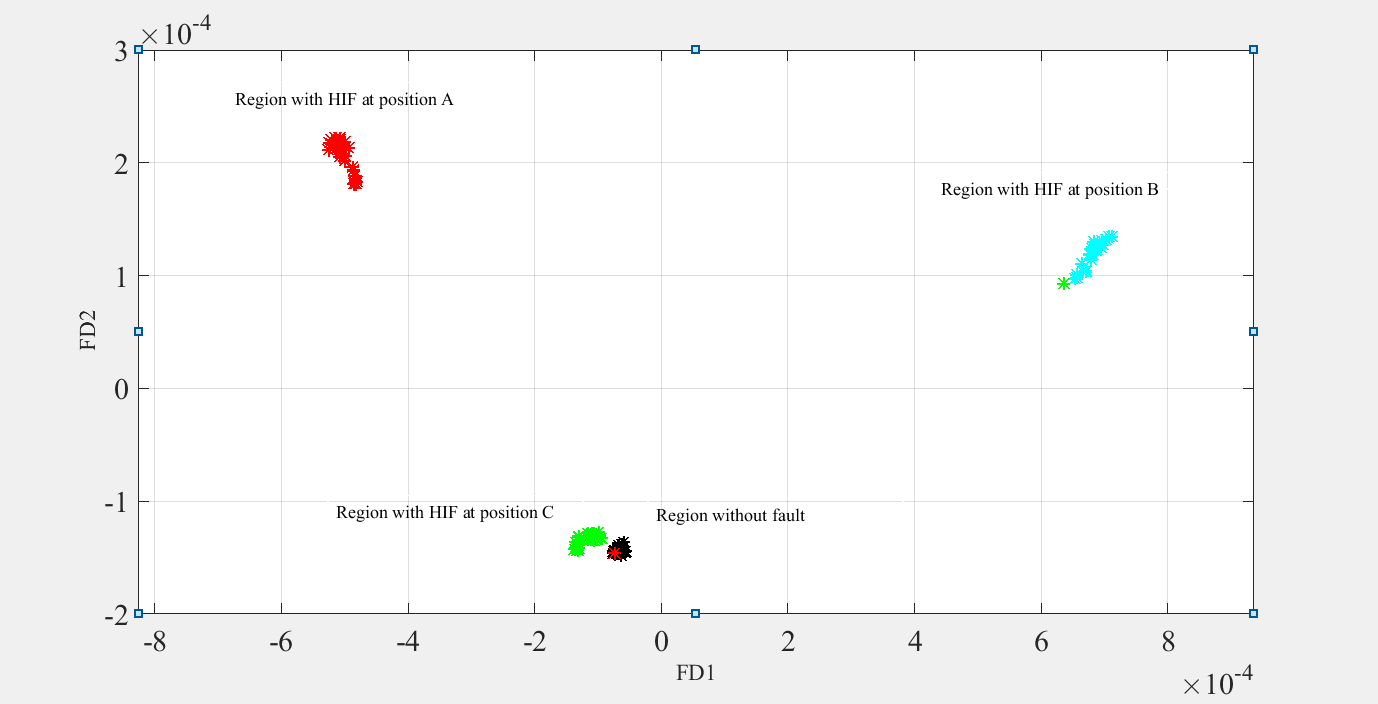}
	\makeatother 
	\caption{{Projection of training data in two dimensional space by FDA}}
	\label{FDAprojection}
\end{figure}

\begin{figure}[tb]
	\centering \includegraphics[height=4.75cm, width=0.5\textwidth]{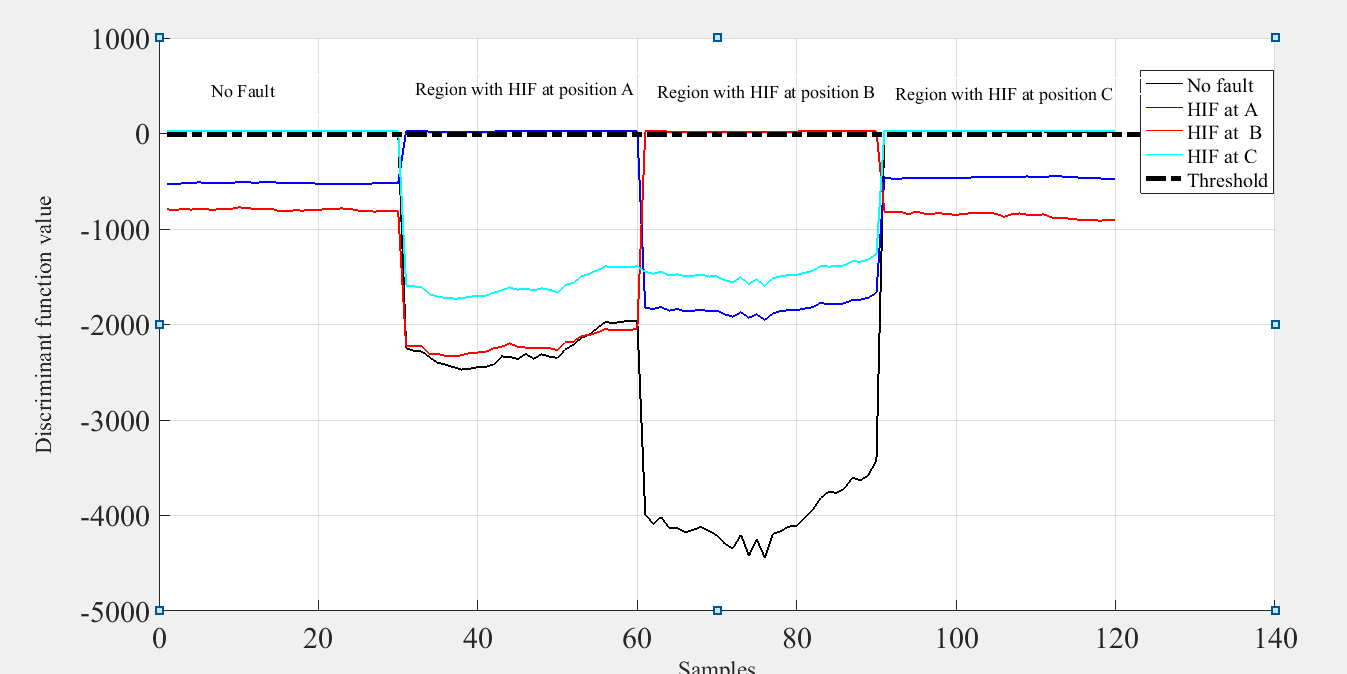}
	\makeatother 
	\caption{{Plot of discriminant function for multiple HIF detection using FDA}}
	\label{discFunPlot}
\end{figure}

\begin{figure}[tb]
	\centering \includegraphics[width=0.5\textwidth]{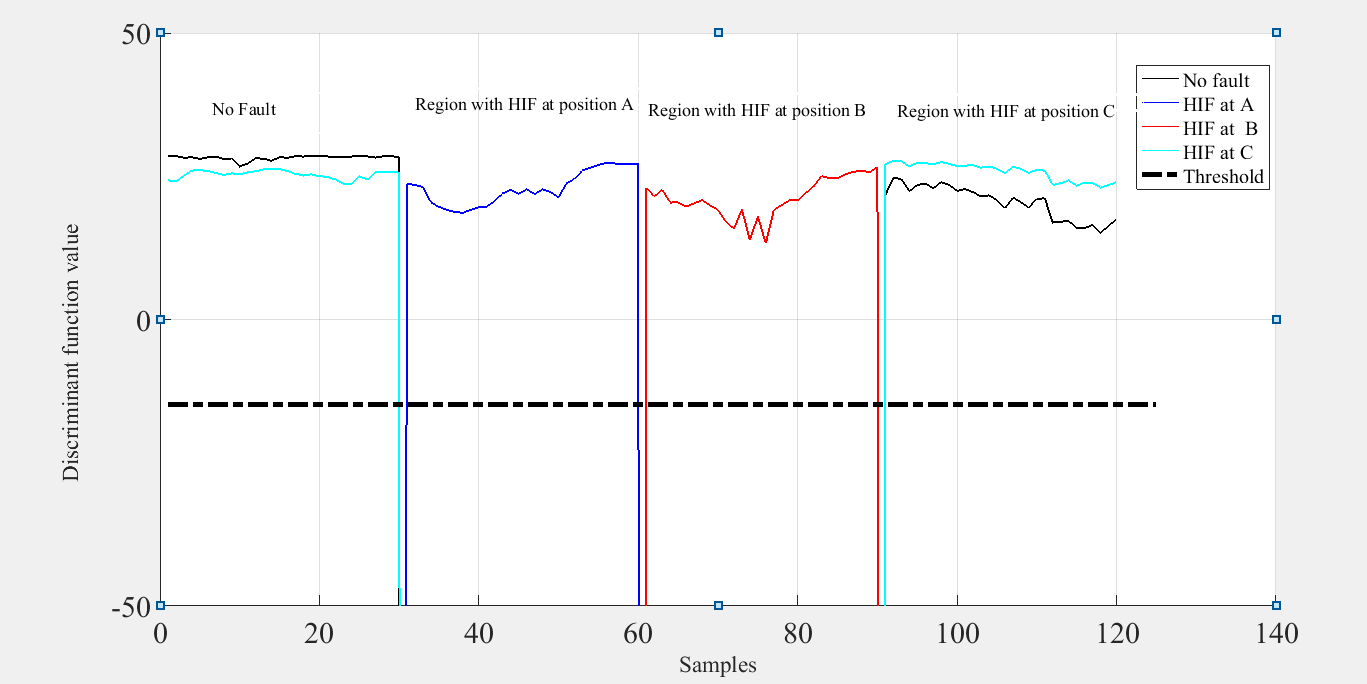}
	\makeatother 
	\caption{{Zoomed view of discriminant function for multiple HIF detection using FDA}}
	\label{discFunZoom}
\end{figure}

FDA is applied for   detection and isolation of high impedance faults in power distribution systems. In order to compute FDA vectors, both normal and faulty data is used, in this work, 60 samples in training data and 40 samples in test data corresponding to each scenario, that is, faulty and non-faulty case. Fig.~\ref{FDAprojection}  shows projection of training data in two-dimensional space by FDA. In second step, after computation of transformation vectors (FDA vectors), the discriminant function is used to test online data. Fig.~\ref{discFunPlot} shows plot of discriminant function in each category. It can be observed that up to first 30 samples, value of discriminant function corresponding to normal case has maximum magnitude, which shows that there is no fault in test feeder. Similarly, after 30 samples, value of discriminant function corresponding to fault at position A has maximum magnitude, which shows that fault at position has occurred. Same is the case with fault at position B and C. Zoomed view of plot is shown in Fig.~\ref{discFunZoom}.

The above results have shown that FDA can successfully isolate/locate HIF. This technique is very well suited for monitoring of power distribution systems.

\subsection{Detection of HIF using SVM}
\begin{figure}
	\centering \includegraphics[width=0.5\textwidth]{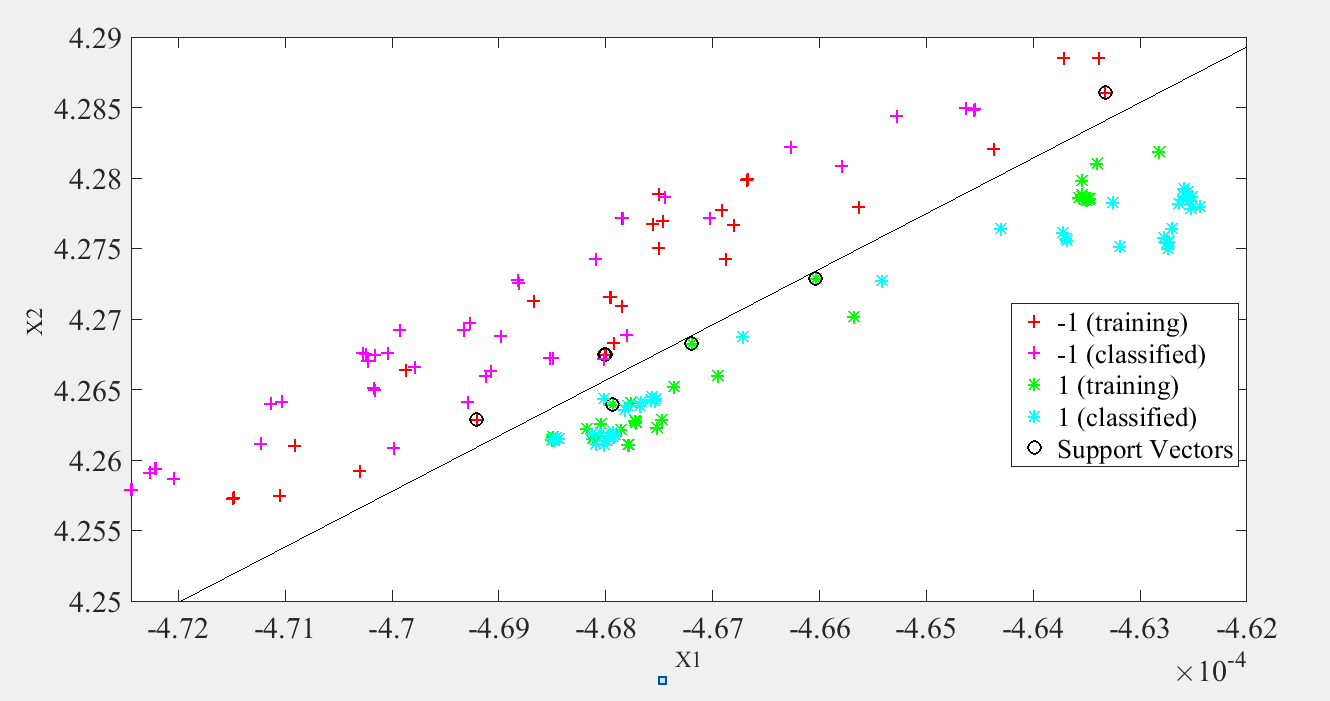}
	\makeatother 
	\caption{{Classification of an HIF using binary-class SVM}}
	\label{binarySVM}
\end{figure}

Support vector machine algorithm has been applied for 29-dimensional data without any dimensional reduction technique. Selection of optimal value of penalty factor is important, this is done by performing nested 3-fold cross validation in original data. With the help of cross validation, average area under the curve was computed for 1000 values of penalty factor between 0.1 and 100. After selection of optimal value of C, SVM classifiers were trained with optimal penalty factor and validated on training data so that generalisation would be checked. Test data of HIF was classified by validated SVM classifier. The predicted labels of test data fairly detects the occurrence of fault, that is -1 for non-faulty data and +1 for faulty data as shown in Fig.~\ref{binarySVM}. 

\subsection{Detection and Classification of HIF using M-SVM}
Binary class SVM can be used for fault detection, but it cannot be used for fault classification. However, in Power distribution systems, discrimination of more than two classes is required, hence, multiclass pattern recognition is often required in monitoring Power distribution systems. Multiclass SVM (M--SVM) classifier is obtained using training of non-fault cases with class label 4, fault at position A with class label 3, fault at position B with class 2, and fault at position C with class 1. In each classifier, during training, a Gaussian Radial Basis Function kernel with a scaling factor, sigma ($\sigma$), of 0.5 and a penalty factor of 10 is used. The tolerance value for Karush-Kuhn-Tucker (KKT) condition for the training of data is taken as 0.001. The value of regularization parameter, lambda ($\lambda$) is 1. Test data is classified using the trained classifiers for 50 observations of each data class and predicted labels were differentiated with known data labels.

\begin{figure}
	\centering \includegraphics[width=0.5\textwidth]{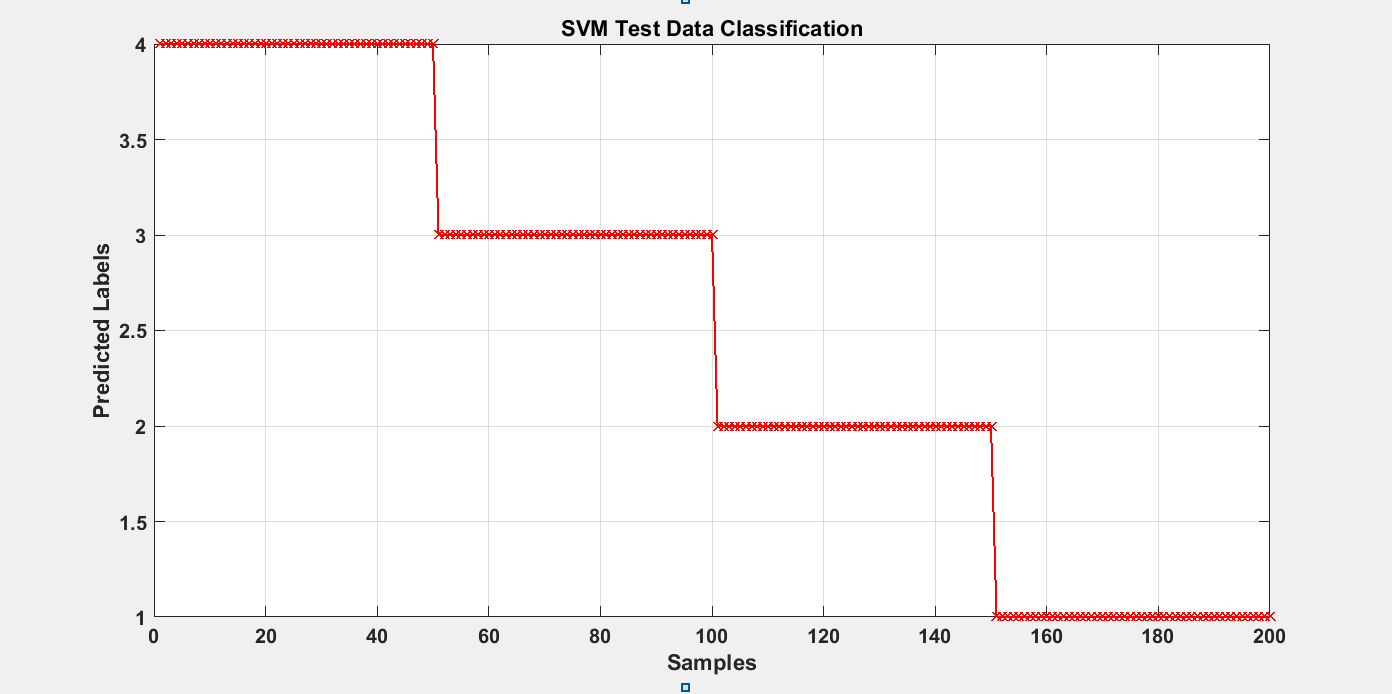}
	\makeatother 
	\caption{{Predicted labels of test data using Multiclass--SVM classifier}}
	\label{predLabelsSVM}
\end{figure}

\begin{figure}
	\centering \includegraphics[height=4.81cm, width=0.5\textwidth]{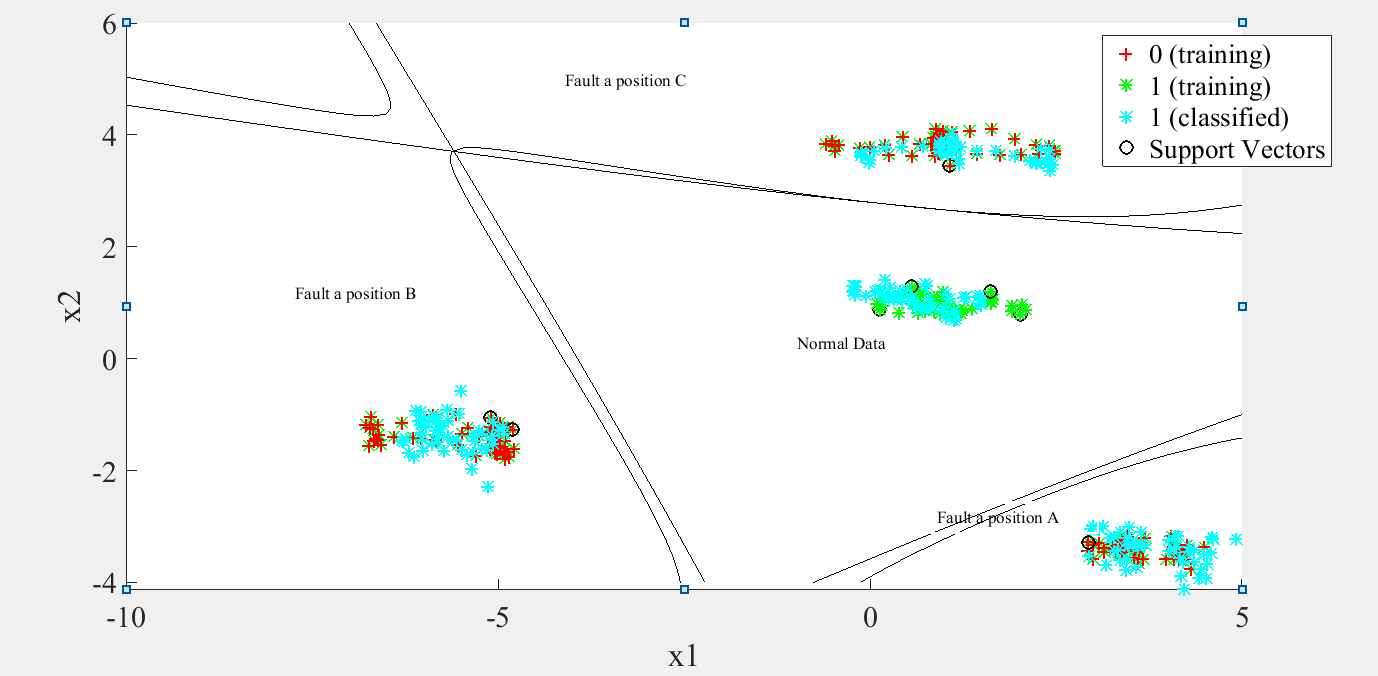}
	\makeatother 
	\caption{{Predicted labels of test data using M--SVM classifier on a 2-D plane}}
	\label{predLabelSVM2d}
\end{figure}

Fig.~\ref{predLabelsSVM} shows that up to 50 samples, the predicted labels belong to normal class data, indicating that there is no fault. After first 50 samples, predicted labels belong to class label 3, indicating that the fault is occurred at position A. After first 100 samples, predicted labels belong to class label 2, indicating that the fault is occurred at position B. similarly, after first 150 samples, predicted labels belong to class label 1, indicating that the fault is occurred at position C. Similarly, Fig.~\ref{predLabelSVM2d} show the score plot of test data for each class of data.

A comparison is presented to evaluate the results of the proposed technique with those from literature and observations have been recorded in Table \ref{tableComp}. After testing the technique on 400 test cases, it is found that proposed method is extremely quick and efficient in detecting HIFs. The proposed method is evaluated through the following performance indices:\\
Dependability: Predicted HIF cases/Actual HIF cases.\\
Security: Predicted non-HIF cases/Actual non-HIF cases.

Table 1 compares the performance indices of the proposed method. It is noted that the proposed method detects all HIF faults under various operating conditions and disturbances. Thus, the proposed method is accurate, reliable and prompt in the detection of High Impedance Faults.

\begin{table*}[]
	\caption{Comparison of performance indices of the proposed M-SVM method with previous techniques}
	\begin{center}
		\begin{tabular}{|c|c|c|}
			\hline
			\textbf{Method} & \textbf{Security (\%)} & \textbf{Dependability (\%)}\\
			\hline
			Wavelet transform \citep{chen2016detection} & 68.5 & 72 \\ \hline
			Time frequency transform \cite{samantaray2008high} & 81.5 & 98.3 \\ \hline
			Morphological gradient \citep{sarlak2011high}  & 96.3 & 98.3 \\ \hline
			Mathematical Morphology \citep{gautam2012detection} & 100 & 100 \\ \hline
			The proposed method (M-SVM) & 100 & 100 \\ \hline
		\end{tabular}
		\label{tab1}
	\end{center}
\label{tableComp}
\end{table*}

It can be seen that M-SVM can easily classify high impedance faults at different locations with load variation and capacitor switching. So, we can conclude that SVM based techniques can successfully detect and  locate HIFs in a Power Distribution Network.

\section{Conclusion}
In this research paper, high impedance fault detection and classification in power distribution systems has been studied using data driven techniques. 
Source-diode-resistance model consisting of two diodes with opposite polarity connected to DC sources is utilised to simulate the high impedance fault. Data driven techniques including PCA, FDA, and SVM are applied to detect/classify HIFs. PCA along with Hotteling's $T^2$ statistics to detect HIFs, it is demonstrated that PCA successfully detects HIF but it cannot classify HIFs. Compared to that, FDA can also successfully classify/locate the fault. Further superior results are achieved by M-SVM, fault classification rate of SVM is better than FDA. M-SVM algorithm can detect all types of HIF and is also robust against capacitor and load switching transients in distribution network.



\section*{Conflict of Interest}
The authors declare no conflict of interest.

\bibliography{ref}

\end{document}